\documentclass[a4paper,pra,twocolumn]{revtex4}
\usepackage{mathptmx}
\usepackage{graphicx}

\newcommand{\dif}{\mathrm{d}}
\newcommand{\unit}[1]{\;\mathrm{#1}}
\newcommand{\degrees}{^{\circ}}

\newcommand{\rb}{$^{87}$Rb}
\newcommand{\rbalt}{$^{85}$Rb}

\newcommand{\fig}[1]{Fig.~\ref{#1}}
\newcommand{\text}[1]{\textrm{#1}}

\begin{document}

\title{A frequency-doubled, pulsed laser system for rubidium manipulation}

\author{J.~Dingjan}
\email[Corresponding author: ]{jos.dingjan@iota.u-psud.fr}
\author{B.~Darqui\'e}
\author{J.~Beugnon}
\author{M.P.A.~Jones}
\author{S.~Bergamini}
\altaffiliation[Now at ]{University College London, UK}
\author{G.~Messin}
\author{A.~Browaeys}
\author{P.~Grangier}
\affiliation{%
  Laboratoire Charles Fabry de l'Institut d'Optique,  Centre Universitaire d'Orsay, France}
%\mail{J. Dingjan, \makeatletter\texttt{jos.dingjan@iota.u-psud.fr}\makeatother}

\date{\today}

\begin{abstract}
We have constructed a pulsed laser system for the manipulation of cold \rb{} atoms. The system combines optical telecommunications components and frequency doubling to generate light at $780\unit{nm}$. Using a fast, fibre-coupled intensity modulator, output from a continuous laser diode is sliced into pulses with a length between $1.3$ and $6.1\unit{ns}$ and a repetition frequency of $5\unit{MHz}$. These pulses are amplified using an erbium-doped fibre amplifier, and frequency-doubled in a periodically poled lithium niobate crystal, yielding a peak power up to $12\unit{W}$. Using the resulting light at $780\unit{nm}$, we demonstrate Rabi oscillations on the $F=2\leftrightarrow F'=3$-transition of a single \rb{} atom.\\[\baselineskip]
\end{abstract}

\maketitle

\section{Introduction}
As part of our experiments on quantum computing using individually trapped, cold rubidium atoms~\cite{sch01a,Pro02a}, we wish to turn a single atom into a single photon source. In order to drive transitions on the D$_2$ line of \rb{}, we require a laser system at $780\unit{nm}$ capable of generating $\pi$-pulses. We seek a system that provides a higher peak power than a diode laser in combination with an intensity modulator, while avoiding the complexity and cost of a titanium-sapphire laser or MOPA system.

It is a happy coincidence that the D$_2$-line of rubidium is almost exactly twice the frequency of one of the standard optical telecommunication frequencies, channel C21 of the Dense Wavelength Division Multiplexing (DWDM)-grid ($f_c=192.10\unit{THz}$)~\cite{ITU}. In the case of \rb{}, the frequency mismatch between twice this frequency and the D$_2$-transition ($f_{\text{D2}}=384.2305\unit{THz}$~\cite{rubidium}) is only about $30\unit{GHz}$, a frequency-difference that, as we will show, can be overcome relatively easily. Therefore, in principle it is possible to build a laser system using optical telecommunication components designed to operate around $\lambda=1560\unit{nm}$, and use frequency doubling to get coherent light at $780\unit{nm}$. This idea has been used previously to stabilise $1560\unit{nm}$ diode lasers on the \rb{} D$_2$-line~\cite{Mah96a,Bru98a}.

This approach offers many advantages. There is a large market for optical telecommunication components, and thus they benefit from a large amount of research and development. Furthermore, optical telecommunication systems are frequently used outside well-controlled laboratory environments, need to be reliable and cost effective, and as a result are rugged and have a high passive stability. And not least of all, because of larger sales volumes, optical telecommunication components are often relatively affordable.

Of course, the flip side of using industry-standard components in a laboratory setting is that the behaviour of components is only characterised insofar as it affects actual telecommunication applications, and that we may end up using components in ways for which they were never designed. As we shall see, examples of the latter include using an Erbium Doped Fibre Amplifier (EDFA) designed for CW operation to amplify short pulses, and operating a diode laser $30\unit{GHz}$ from its design working point.

\section{Description of the system}
\subsection{Requirements}
As mentioned in the introduction, we wish to generate $\pi$-pulses to couple either of the two ground state hyperfine levels of \rb{} to, primarily, the $F'=2$ and $F'=3$ hyperfine levels of the $5^2P_{3/2}$ excited state. From this follow several requirements on our laser system.

The pulses generated by this system should be shorter than the spontaneous lifetime of the excited state, $t_{\text{sp}}=26.2\unit{ns}$~\cite{rubidium}. At the same time, to ensure state selectivity, they should be long enough that their bandwidth remains smaller than the frequency separation between the $F'=2$ and $F'=3$ excited state hyperfine levels ($\Delta \nu_{23} = 267\unit{MHz}$~\cite{rubidium}). This imposes a best-case lower limit (for Fourier-limited Gaussian pulses) on the full-width, half-max (FWHM) pulse duration of $\tau>1.6\unit{ns}$. From these two requirements, we set our aims at pulse lengths between $\sim 2$ and $\sim 6\unit{ns}$.

An upper bound on the repetition frequency of the laser comes from the requirement that the pulse period be much longer than the spontaneous lifetime. This ensures that the atom will, with near certainty, have relaxed to the ground state before the arrival of the next excitation pulse. A lower bound on the repetition frequency follows from the wish to maximise the rate of photon emission, both from a fundamental and an experimental point of view. For this system, we have settled on a pulse period of $200\unit{ns}$.

Furthermore, as we wish to couple both the ground state hyperfine levels, the central laser frequency must be tunable over at least the separation between the $F=1$ and $F=2$ ground states of \rb{}, $\Delta\nu_{12}=6.83\unit{GHz}$~\cite{rubidium}. In addition to giving us a greater flexibility in quantum-optical experiments, such a tunability permits easy diagnostics, as it allows us to do quick scans of rubidium spectra in a vapour cell.

Finally, to have a reasonable power budget, we require a peak power of at least $1\unit{W}$. This will allow us to achieve $\pi$-pulses without having to focus our excitation beam to more than $\sim 1\unit{mm}$, greatly simplifying the alignment of the excitation beam onto the trapped atom.

\subsection{Implementation}
\begin{figure}
  \includegraphics[scale=0.65]{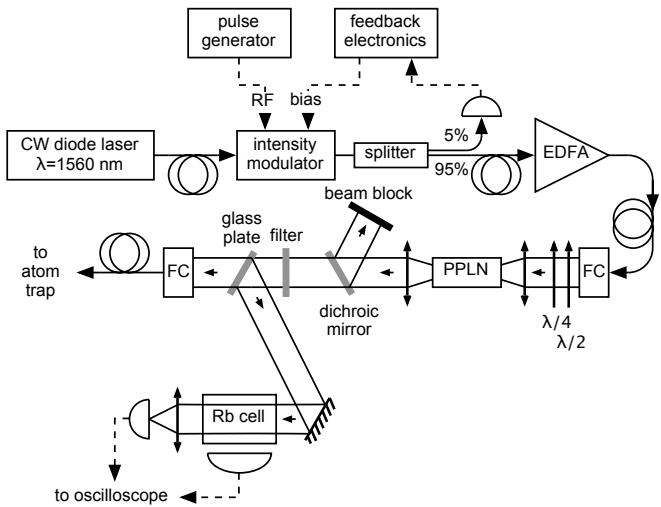}
  \caption{\label{fig:lasersystem}Schematic of the laser system. CW light from a diode laser at $1560\unit{nm}$ is sliced into pulses using an intensity modulator. These pulses are amplified using an erbium-doped fibre amplifier (EDFA) to an average power of $0.8\unit{W}$. After frequency-doubling in a periodically-poled lithium niobate (PPLN), we obtain pulses at $780\unit{nm}$ with an average power of $80\unit{mW}$. A small fraction of this light is sent to a rubidium cell to verify laser tuning. The rest of the light is coupled into a single mode fibre and transmitted to the experimental setup. FC: Fibre Coupler.}
\end{figure}

Using the ideas mentioned in the introduction, and roughly following  the ideas for a laser source at $532\unit{nm}$~\cite{Pop00a,Bev02a}, we have constructed a laser system that meets the specifications outlined in the previous section. In \fig{fig:lasersystem}, we give an overview of this system. In the following, we mention specific components for reference only. We have not made an exhaustive search of all available alternatives, and similar components from other manufacturers may give similar results.

At the start of the chain, we have a JDS-Uniphase CQF935/808 series continuous-wave, distributed feedback (DFB) diode laser in a butterfly package, with a designed output power of $50\unit{mW}$. These lasers are available with design wavelengths (using the operating current and temperature as specified by the manufacturer) between $1527$ and $1610\unit{nm}$, with $0.4$ or $0.8\unit{nm}$ steps, and a linewidth of $< 1\unit{MHz}$. Our particular laser has a design wavelength of $1560.61\unit{nm}$ for $50\unit{mW}$ output power, at a drive current of $\sim 338\unit{mA}$ and a diode temperature of $34.8\unit{\degrees C}$. We operate this laser at an output power of $30\unit{mW}$, so as not to exceed the maximum power tolerance of other components. This has the side effect of shifting the lasing wavelength towards the blue, even more than is needed to bridge the above-mentioned $\sim30\unit{GHz}$ gap between twice the channel frequency $2 f_c$ and the \rb{} transition frequency $f_{\text{D2}}$. To get back to the required laser frequency, the diode temperature has to be increased to $36.5\unit{\degrees C}$.

By slowly modulating either the diode temperature or the drive current, the laser frequency can be tuned over $>4\unit{GHz}$, without mode-hopping. For our specific diode laser, we have measured the dependence of laser frequency on diode temperature or drive current to be $\dif \nu/\dif T = -11 \unit{GHz/\degrees C}$ and $\dif \nu/\dif I= -0.20\unit{GHz/mA}$ (measured at $\lambda=1560\unit{nm}$).

The CW output of this laser is sliced into pulses with a width of $1.3$--$6.1\unit{ns}$ and a repetition frequency of $5\unit{MHz}$ by a JDS-Uniphase 100219-series fibre-optic, chirp-free intensity modulator. Internally, this intensity modulator takes the shape of a Mach-Zehnder interferometer, with balanced lithium-niobate electro-optic phase shifters in both interferometer arms. This modulator is biased at zero transmission using a lock-in feedback system. On the bias input, the voltage that is needed to go from minimum to maximum transmission $V_{\pi,bias}\approx7.9\unit{V}$, while for the RF input $V_{\pi,\text{RF}}\sim 4\unit{V}$. The specified extinction ratio of the modulator is $\geq 20\unit{dB}$.

The RF input of the intensity modulator is driven from an AVTech AV-1-C pulse generator, triggered by an external master clock. This external clock provides a more stable $5\unit{MHz}$ repetition frequency than the internal clock of the pulse generator, as well as more synchronisation flexibility.

After the intensity modulator, a fibre splitter sends 5\% of the transmitted light to a photodiode, which provides the feedback signal for a lock-in servo loop to bias the intensity modulator at zero transmission.

The next stage in the chain is a Keopsys KPS-BT-C-30-PB-FA high power C-band erbium-doped fibre amplifier with pre-amplifier and booster. It is designed to amplify CW input. For CW input powers $> 0.1 \unit{mW}$ ($-10\unit{dBm}$), the output power saturates at $1\unit{W}$ ($30\unit{dBm}$), whereas the CW small signal gain is $> 55\unit{dB}$. The maximum input power is of the order of  several mW, above which the thermal protection will switch off the device.

However, provided the average input power does not exceed the maximum CW input power, the amplifier will amplify pulsed input light, without measurably degrading pulse shapes or widths, to similar \emph{average} output power levels as for CW input. As a result, after the fibre amplifier we obtain light pulses with an average power of $\sim 0.8\unit{W}$. With a duty cycle between $1/33$ and $1/150$, this gives us a peak power of $26$--$120\unit{W}$. 

This high power leads to polarisation effects in the $20\unit{cm}$ patch fibre used to couple light from the amplifier to a fibre coupler. After the fibre coupler, the light is polarised elliptically, with the ellipticity and orientation of the polarisation ellipse strongly dependent on the average power. Using a half-wave plate and a quarter-wave plate, the light is restored to linear polarisation, with a purity $>99\%$. After the initial warmup of the system (most notably the amplifier), the optimum setting for these two wave plates is stable, both on a short timescale (minutes) and on a long timescale (days and weeks).

The peak power we obtain after the fibre amplifier is sufficient to achieve a typical single-pass frequency-doubling efficiency of up to $15\%$ in a periodically-poled lithium niobate (PPLN) crystal with a length $40\unit{mm}$ obtained from HC Photonics. The doubling bandwidth of the nonlinear crystal exceeds the relevant frequency scale in our experiments, the separation between the $F=1$ and $F=2$ ground state hyperfine levels. The PPLN crystal is temperature-tuned to optimise the frequency-doubling efficiency, with a typical operating temperature around $200\unit{\degrees C}$. To this end, the crystal is placed in an oven that keeps the temperature constant to within $0.1\unit{\degrees C}$. For optimum efficiency, the free-space beam at $1560\unit{nm}$ is focused into the crystal, with the Rayleigh range inside the medium equal to $20\unit{mm}$ to match the length of the crystal.

After the crystal we recollimate the beam, and filter out the transmitted light at $1560\unit{nm}$ using a dichroic mirror with a transmissivity of $>85\%$ at $780\unit{nm}$ and a reflectivity of $>99.5\%$ at $1560\unit{nm}$, and a band-pass filter that transmits $>85\%$ at $780\unit{nm}$ and that has an optical density of OD4 at $1560\unit{nm}$. After this, we are left with a beam of light at $780\unit{nm}$, with an average power of typically $80\unit{mW}$ and a peak power of $2.6$--$12\unit{W}$. Most of this light is coupled into a single-mode fibre, to transport it to a different optical table. As we achieve a coupling efficiency into the fibre of $\sim 90\%$, we conclude that the beam at $780\unit{nm}$ is very close to Gaussian.

The remaining fraction of light is passed through a rubidium vapour cell, to monitor the tuning of the source laser. Fluorescence from this cell is collected using a slow photodiode (too slow to see the individual pulses). The passive frequency stability of the system is high; over the course of a day, the laser frequency drifts less than $\sim 50\unit{MHz}$, and even after switching off the laser at night, and switching it back on the next day, it will return to within $<50\unit{MHz}$ of the previously set frequency.

\section{Diagnostic measurements}
\begin{figure}
  \includegraphics[scale=0.95]{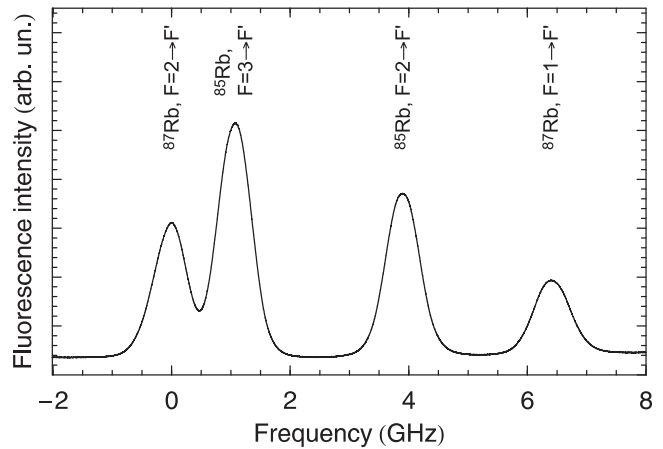}
  \caption{\label{fig:rbspectrum}Fluorescence signal from a rubidium vapour cell versus frequency (averaged over 1000 runs).}
\end{figure}

To demonstrate the broad tunability of this source, we apply a slow current modulation to the laser system, and measure the fluorescence intensity from the rubidium cell. In \fig{fig:rbspectrum} we plot this fluorescence signal versus frequency. As we can see we observe four broad maxima in the amount of emitted fluorescence, corresponding to transitions from the two ground state hyperfine levels of either isotope \rb{} and \rbalt{}. Because of Doppler broadening in the (thermal) rubidium vapour, the hyperfine levels of the excited state cannot be resolved. By tuning the diode laser's temperature and drive current, its frequency can be centred on the target transition, with an accuracy $<30\unit{MHz}$.

\begin{figure}
  \includegraphics[scale=0.95]{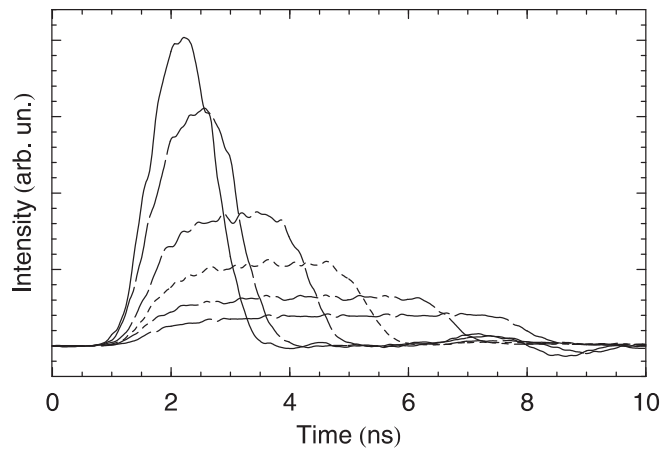}
  \caption{\label{fig:pulseshapes}Pulse shapes for a full width at half maximum (FWHM) pulse duration of $1.3$, $1.6$, $2.6$, $3.6$, $5.0$, and $6.1\unit{ns}$, respectively.}
\end{figure}

Furthermore, using a fast photodiode we have measured the pulse shape for various pulse lengths between $1.3\unit{ns}$ and $6.1\unit{ns}$, as plotted in \fig{fig:pulseshapes}. These pulse shapes are compatible with square pulses convolved with the impulse response of the photodiode and detection electronics, which we have measured to be approximately gaussian with a width of $0.9\unit{ns}$. Between the pulses, the level of residual light is $\sim 0.1$--$0.5\%$ of the pulse maximum, depending on the pulse length. For pulses of $4\unit{ns}$, the residual light level is $\sim 0.3\%$. Most of this light is actually due to the modulation signal on the bias of the intensity modulator, and is an unavoidable side effect of the feedback loop for the modulator. 
 % Dataset of 16 Nov 2004

Much more important to us is the spectral width of the delivered pulses, or the time-bandwidth product. Since the laser system will be used on rubidium atoms, the relevant frequency scale is set by the separation between the $5^2\mathrm{P}_{3/2}$ excited state hyperfine levels of \rb{}. Our target state is the $F'=3$ state, so we will compare the spectral width of the laser pulses with the $267\unit{MHz}$ separation between $F'=2$ and $F'=3$.

\begin{figure}
  \includegraphics[scale=0.92]{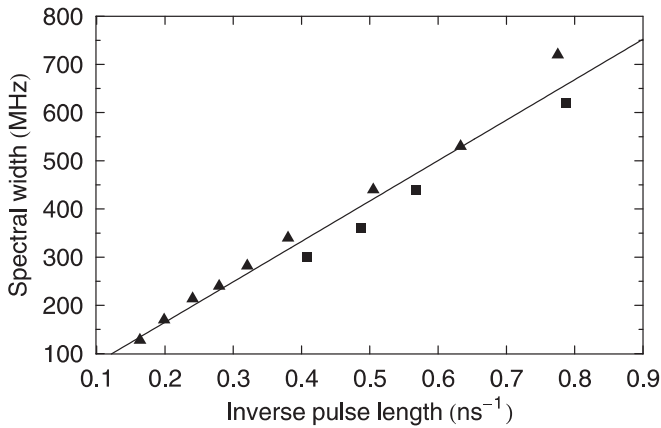}
  \caption{\label{fig:timebandwidth}FWHM spectral width as a function of the inverse of the FWHM pulse duration, measured directly after the fibre amplifier (squares), and after the nonlinear crystal (triangles). Straight line is a fit to both datasets, with a slope of $0.84\unit{GHz\times ns}$.}
\end{figure}

In \fig{fig:timebandwidth} we plot the full width at half maximum (FWHM) spectral width of our pulses as a function of the FWHM pulse duration, measured with a Fabry-Perot interferometer either directly after the fibre amplifier, or after the nonlinear crystal. We find that the spectral width is proportional to the inverse of the pulse duration, with a (fitted) time-bandwidth product of $0.84(5)$, compared to $0.89$ for rectangular pulses. From this, and the measured temporal pulse shapes, we conclude that the pulses are Fourier-limited. For pulses that are longer than about $3\unit{ns}$ the spectral width is less than the separation of the $F'=2$ and $F'=3$ excited states. This allows us to individually address the excited state hyperfine levels.

\section{Application: Rabi oscillations}
To demonstrate the viability of this laser system for actual cold atom experiments, we have used its output to drive Rabi-oscillations on the $F=2$--$F'=3$-transition of \rb{}, and to perform well-controlled optical $\pi$-pulses on this transition.

As described in references~\cite{sch01a,sch02a}, we trap individual \rb{}-atoms in an optical dipole trap formed by focusing light from a diode laser at $810\unit{nm}$ with a high-numerical-aperture objective. The trapping volume is of the order of $1\unit{\mu m^3}$. Because of this small size, we either trap 0 or 1 atom, and never more. We illuminate the trapped atom with the pulsed laser system described in this article, and collect fluorescence from the trapped atom and image it onto a single-photon APD.

To show that we can drive Rabi-oscillations, we vary the intensity of the laser pulses while keeping the pulse length and detuning constant. Since the duration of the pulses ($4\unit{ns}$) is considerably smaller than the spontaneous lifetime of rubidium ($26.24\unit{ns}$), we expect most of the fluorescence light to be emitted \emph{after} the probe pulses, rather than during. At the same time, the time between probe pulses ($200\unit{ns}$) is so large that the atom will have decayed to the ground state before a second probe pulse arrives. In that case, the number of photons emitted per second is proportional to the excited state occupation of the atom at the end of the pulse. As the Rabi-frequency is proportional to the square root of the intensity, we expect to see, for constant pulse length, a periodic dependence of the photon emission rate on the square root of the probe intensity.

\begin{figure}
  \includegraphics[scale=0.52]{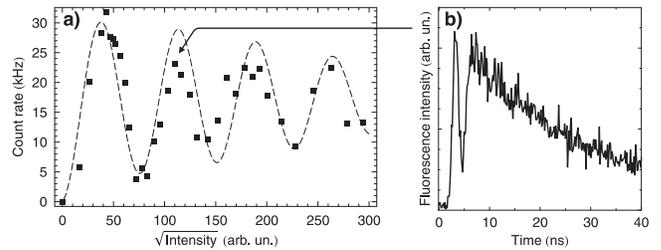}
  \caption{\label{fig:rabi}a) Photon count rate versus the square root of the probe power. The dashed line is a theoretical curve for a two-level model with $10\%$ intensity fluctuations. b) Fluorescence signal versus time, for excitation using $3\pi$-pulses. We clearly see that during the pulse, the atom makes one and a half Rabi oscillation.}
\end{figure}

In \fig{fig:rabi}a we plot the photon count rate as a function of the square root of the probe power, while in \fig{fig:rabi}b we show a time-resolved fluorescence signal in the case of $3\pi$-pulses. In the left plot, we clearly see four periods of Rabi-oscillations. Taking into account the collection efficiency of the imaging system, we find that the maximum excitation probability of our atom is $(95\pm5)\%$. In addition, we see that for $2\pi$-pulses (and multiples thereof), the excitation probability does not descend to zero, but stays at a finite value. This reflects the finite probability of emitting a photon \emph{during} the excitation pulse. In Figure~\ref{fig:rabi}b, we observe one and a half Rabi oscillation, followed by free exponential decay. These results, as well as the use of this laser system to generate single photons, are published elsewhere~\cite{Dar05a}.

For increasing power we see that the visibility of the Rabi oscillations decreases. This is due to fluctuations in the power of the excitation laser, leading to a smearing out of the Rabi oscillations.  The dashed line in Figure~\ref{fig:rabi}a is a theoretical calculation of Rabi oscillations as could be observed for a two-level system in the presence of $10\%$ intensity fluctuations in the exciting pulses. These intensity fluctuations have their origin in the modulation signal for the feedback loop around the intensity modulator, and we expect they can be reduced by redesigning the feedback electronics. It may even be possible to completely do away with the feedback loop, and tune the intensity modulator to minimum transmission ``by hand''. We note that these intensity fluctuations hardly affect $\pi$-pulses, as shown by the above-mentioned excitation probability of $\sim95\%$.

\section{Conclusion}
In conclusion, we have constructed a laser system that is capable of generating laser pulses with a wavelength of $780\unit{nm}$, a pulse duration between $1.3$ and $6.1\unit{ns}$, and a peak power of $2.6$--$12\unit{W}$. The system uses optical telecommunications components with a design wavelength of $1560\unit{nm}$ and a frequency-doubling nonlinear crystal. As a first demonstration of the use of this system, we have shown that we can perform up to $4$ Rabi oscillations.

If we choose our parameters in such a way that during a pulse we perform exactly $1/2$ a Rabi oscillation, we have a single-shot coherent excitation of our atoms, with a very high excitation probability ($\sim95\%$).

In the near future we plan to extend this laser system, by adding a phase modulator between the laser diode and the intensity modulator. This will allow us to generate frequency-chirped laser pulses, which will be used to perform rapid adiabatic passage (RAP) on the same optical transition used for the above-mentioned Rabi oscillations. We expect that in this way we can use shorter excitation pulses, without losing level selectivity.

\begin{acknowledgments}
We thank Patrick Georges for his assistance in designing the pulsed laser system, and Andr{\'e} Villing and Fr{\'e}d{\'e}ric Moron for indispensable electronics support. This work was supported by the European Union through the Research Training Network ``CONQUEST'', and the IST/FET/QIPC project "QGATES". M.~Jones was supported by EU Marie Curie Fellowship MEIF-CT-2004-009819.
\end{acknowledgments}

%\bibliography{articles}

\end{document}